\documentclass[prb, reprint, twocolumn]{revtex4}
\usepackage{float}
\usepackage{amsmath}
\usepackage{amsfonts}
\usepackage{graphicx}
\usepackage{hyperref}

\begin{document}

\title{High-accuracy longitudinal position measurement using self-accelerating light}
\author{Shashi Prabhakar}\email{shashi.sinha@tuni.fi}
\author{Stephen Plachta}
\author{Marco Ornigotti}
\author{Robert Fickler}
\affiliation{Photonics Laboratory, Physics Unit, Tampere University, Tampere, FI-33720, Finland}

\date{\today}

\begin{abstract}
Radially self-accelerating light exhibits an intensity pattern that describes a spiraling trajectory around the optical axis as the beam propagates. 
In this article, we show in simulation and experiment how such beams can be used to perform a high-accuracy distance measurement with respect to a reference using simple off-axis intensity detection.
We demonstrate that generating beams whose intensity pattern simultaneously spirals with fast and slow rotation components enables a distance measurement with high accuracy over a broad range, using the high and low rotation frequency, respectively.
In our experiment, we achieve an accuracy of around 2~$\mu$m over a longitudinal range of more than 2~mm using a single beam and only two quadrant detectors.
As our method relies on single-beam interference and only requires a static generation and simple intensity measurements, it is intrinsically stable and might find applications in high-speed measurements of longitudinal position.
\end{abstract}

\maketitle

\section{Introduction} \label{sec:intro}
Structuring the spatial shape of light fields has become a broad research field spanning areas from the foundations of optics to optical communication, materials processing, quantum optics, and microscopy, to name a few \cite{rubinsztein2016roadmap}.
Amongst many interesting features structured light may have, one in particular has attracted a lot of attention and might even be seen as the starting point of the field, namely the azimuthal phase structure connected to the orbital angular momentum (OAM) of light \cite{padgett2017orbital}.
Light fields carrying such OAM have a transverse phase of the form $\exp(-i\ell\varphi)$, where $\varphi$ is the azimuthal coordinate and $\ell$ defines the quanta of OAM each photon carries \cite{allen1992orbital}. 
These transverse scalar modes are commonly known as vortex modes, or donut beams as they have a phase singularity and, thus, an intensity null along the optical axis.
Over the last decades, various techniques to imprint such twisted structures have been established, e.g., spiral phase plates \cite{beijersbergen1994helical}; holographic generation using spatial light modulators \cite{heckenberg1992generation, carpentier2008making, forbes2016creation}; cylindrical lenses \cite{beijersbergen1993astigmatic}; $q$- and $J$-plates \cite{marrucci2006optical,larocque2016arbitrary, devlin2017arbitrary}; or direct generation of the light field inside the cavity of a laser \cite{forbes2019structured}.

One particular family of modes, whose higher orders have an azimuthal phase ramp, is known as Bessel beams. 
Bessel beams are propagation invariant light fields described by Bessel functions \cite{gori1987bessel}. 
They have received significant attention due to being diffraction-less \cite{zahid1989directionality, vetter2019realization} and self-healing \cite{chu2012analytical, mcgloin2005bessel}. 
While zero-order Bessel beams have an intensity maximum along the optical axis and do not carry OAM, higher orders have a twisted phase structure leading to well-defined quanta of OAM per photon.
In addition, Bessel beams are comparatively easy to generate using a ring aperture, with or without an azimuthally varying phase, at the back focal plane of a lens ($k$-space).
The light fields in the focal plane of the lens, which have undergone an optical Fourier transformation, then resemble the theoretical Bessel beams; however, they have a finite beam extent and are therefore called Bessel-Gauss beams. 
If not only one ring but multiple rings of different radii and different orders are put into the back focal plane of the lens, a superposition of multiple Bessel beams is generated.
Interestingly, the obtained superposition structures show the peculiar feature of spiraling around the optical axis along the propagation direction if the constituents forming the superposition have different OAM values \cite{chavez1996nondiffracting, schechner1996wave, abramochkin1997generation, paakkonen1998rotating}.
This property of complex superpositions of higher-order Bessel beams has been the focus of various research efforts.
Thorough theoretical and experimental studies of such spiraling beams have been performed, which have been enabled by the progress in experimental techniques of generating such beams with high precision and flexibility \cite{tervo2001rotating, kotlyar2007rotation, vasilyeu2009generating, vetter2015optimization, schulze2015accelerated}.

Contrary to Airy beams, whose self-accelerating character is essentially given by the fact that an observer in a reference frame solidal with an Airy beam would experience a tangential fictitious force \cite{greenberger1980comment} that results in their characteristic parabolic propagation profile, radially self-accelerating beams (RSABs) are characterised by a centrifugal fictitious force linked to their characteristic spiraling motion \cite{vetter2014generalized}.
As such they are also distinct from another class of self-accelerating beams, recently investigated in \cite{vetter2017real, webster2017radially}.

In this work, we demonstrate a novel application of spiraling beams as a means to determine the longitudinal distance with high accuracy by only measuring the intensity using a quadrant detector, i.e., in a limited number of off-axis locations.
First, we briefly introduce the theory behind spiraling beams and describe how superpositions of three Bessel modes can lead to a complex rotating pattern having both quickly and slowly rotating parts at the same time. 
We then show in simulations and experiments that using such a spiraling beam enables the accurate determination of distance over a long range when the intensity using two quadrant detectors (or in minimally three off-axis positions) is recorded.
In the experimental implementation, we are able to achieve an accuracy of around 2~$\mu$m over a range of 2~mm.
The obtained result is mainly limited by the aperture of our optical system, as well as the resolution of the generating and detecting devices.
Hence, the proposed and demonstrated method of measuring a longitudinal distance using structured light might find promising applications, similar to self-accelerating Airy beams, which for example have been used to resolve depth in microscopy applications recently \cite{jia2014isotropic, he2019depth}.
Due to its simplicity and high accuracy, our method nicely complements other available techniques using light, e.g., time-of-flights measurements as used in LIDAR systems \cite{jarvis1983laser}, interferometric approaches \cite{kubota1987interferometer}, or schemes that rely on complex scattering of structured light fields \cite{berg2020microsphere}, to name a few.

\section{Theoretical background} \label{sec:theory}

\subsection{Spiraling light fields}

\begin{figure}[h]
    \centering
    \includegraphics[width=0.4\textwidth]{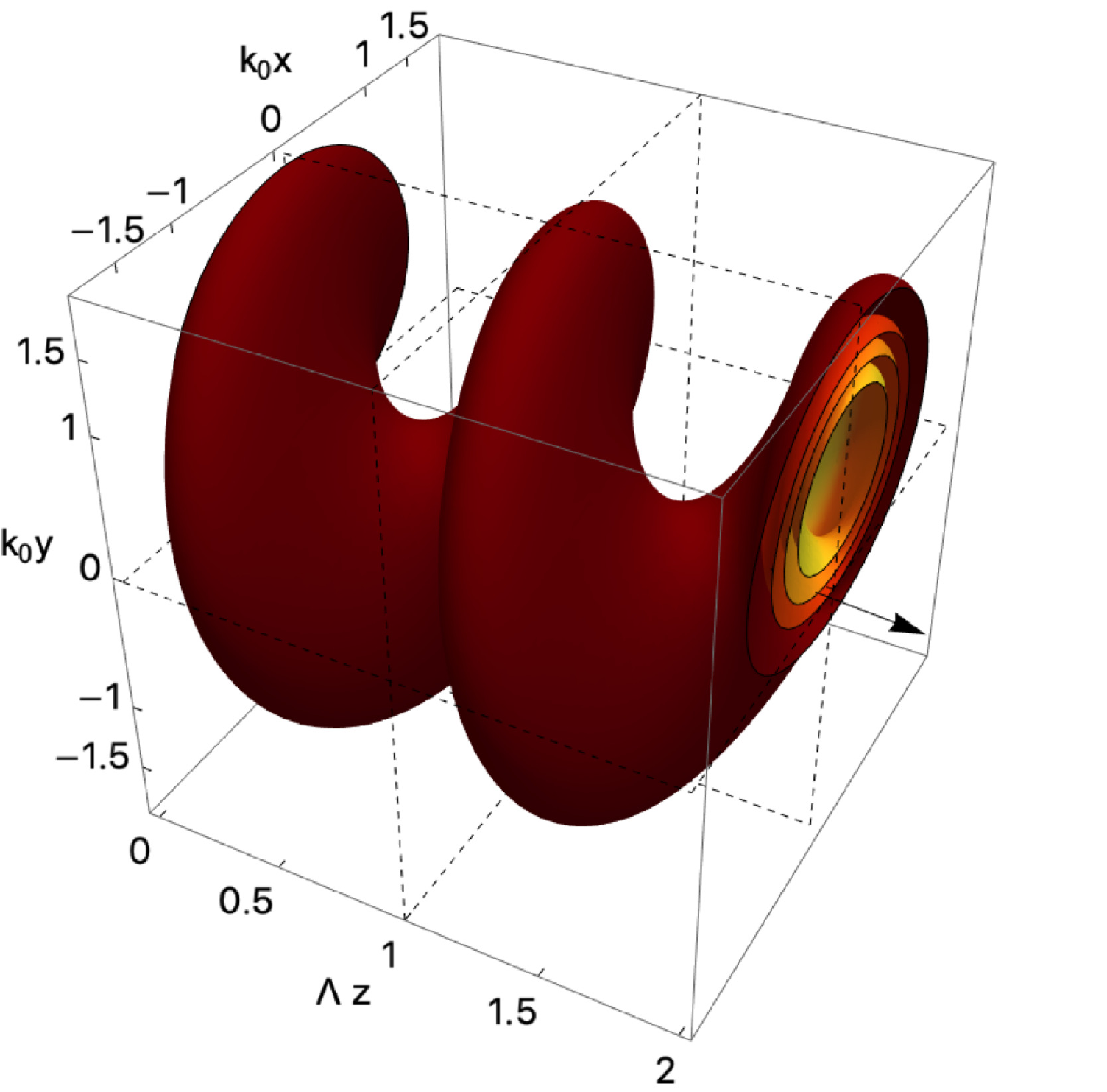}
    \caption{Normalised three-dimensional representation of the propagation of the central lobe of a radially self-accelerating beam along the $z$ direction, as defined in \eqref{eq:field2mo}, consisting of the superposition of $\ell_1=0$ and $\ell_2=1$ Bessel modes, which matches the form of the single-frequency beam used in the experiment. The transverse directions $\{k_0x,k_0y\}$ are normalised to the central $k$-vector of the beam, i.e., $k_0$, while the longitudinal direction (i.e., the propagation direction $z$) has been normalised to the rotation period $\Lambda=2\pi/\omega$, with $\omega$ being the rotation speed defined by \eqref{eq:rotationspeed}. The plot shows propagation up to the first two periods. The colour scale in the picture represents different iso-intensity surfaces, with brighter colours indicating regions of higher intensity. As can be seen, the whole intensity distribution rigidly rotates around the $z$ axis with rotation speed $2\pi/\Lambda$. This peculiar propagation pattern is the result of interference between the various Bessel beam components constituting the radially self-accelerating beam, as described by \eqref{eq:field2mo}.} \label{fig:Theory}
\end{figure}

Measuring the longitudinal position, i.e., a certain distance with respect to a fixed reference, using an intensity structure that changes over propagation requires a light field with well-defined propagation dynamics. 
The recently demonstrated radially self-accelerating, or spiraling, light fields, which show a constant rotation of the intensity pattern along the propagation direction, are a very convenient solution to use. 
While light with more complex propagation dynamics would require sophisticated evaluation procedures, rotating structures allow the determination of the longitudinal position through simple measurements of the rotation angle.
Importantly, such light fields can be easily realized by superimposing (at least) two vortex beams, each having a different OAM value $\ell$ as well as two different longitudinal wave vectors $k_z$ defining the propagation dynamics \cite{vetter2014generalized}. 
The two-component solution, i.e., the so-called helicon beams, can then be written as
\begin{equation}
    u(r,\phi,z) = A_{\ell_1}(r)\exp{[i(k_{1z}z+\ell_1\varphi)]}+ A_{\ell_2}(r)\exp{[i(k_{2z}z+\ell_2\varphi)]},
    \label{eq:field2mo}
\end{equation}
where the indices $1,2$ label the two constituents and $A_\ell(r)$ is a radially-dependent envelope function.
The resulting intensity, $I=u(r,\varphi,z)u(r,\varphi,z)^*$, can then be obtained to be 
\begin{equation}\label{intensity}
    I(r,\varphi,z) \propto \cos^2{[\Delta k z + \Delta \ell \varphi]},
\end{equation}
where  $\Delta k= (k_{1z} - k_{2z})/2$ is the difference between the wave vectors of the two beams and  $\Delta \ell = (\ell_1 - \ell_2)/2$ is the difference between their OAM values. 
Notice, moreover, that we have only kept the part of the intensity that is of importance, i.e., the one including the required angular and $z$-dependence.
For more details we refer the interested reader to earlier works \cite{vetter2014generalized}.
From \eqref{intensity} we find that the angular orientation of the intensity profile $\phi(z)$ changes along the beam propagation according to the relation
\begin{equation}
    \phi(z) = \frac{z \Delta k}{\Delta \ell},
    \label{eq:rot}
\end{equation}
from which the angular velocity can be calculated as    
\begin{equation}
\omega=\frac{\partial\phi(z)}{\partial z}=\frac{\Delta k}{\Delta \ell}. \label{eq:rotationspeed}
\end{equation}
One such beam propagation is shown in Fig. \ref{fig:Theory}, where the spiraling of the mode along the propagation axis is depicted.

In other words, if we measure the intensity over a certain angular region, the intensity along the beam propagation follows a periodic $\cos^2$-function and can be used to determine the longitudinal distance unambiguously within half a period. 
In principle, a simple measurement of the intensity at an off-axis transverse position therefore allows the determination of a distance with arbitrary precision. 
In practical situations, however, errors induced by the generation or measurement of the structure result in an uncertainty in the determination of the intensity's angular position, which leads to a limitation of the longitudinal accuracy.
One direct way to improve the measurement accuracy despite these imperfections is to increase the rotation frequency of the structure with respect to its propagation.
The faster the rotation, the better the accuracy in measuring the longitudinal distance $z$.
Hence, one aim in high-accuracy distance measurement using self-accelerating light fields is to achieve the largest possible difference in the longitudinal wave vectors $\Delta k$ allowed by the optical system.
We also see that the difference in the OAM values $\Delta \ell$ should be kept as small as possible, i.e., the $\ell$-values of the two constituent beams should only differ by 1.

Obviously, improving the longitudinal accuracy by increasing the rotation frequency comes at the cost of a reduced longitudinal range over which an unambiguous determination of the angular position is possible by only examining the intensity pattern.
To circumvent this limitation, it is possible to realize a more complex structure, which shows a rotating intensity pattern that includes two (or more) well-defined rotation frequencies. 
Ideally, the intensity structure should have one very high rotation frequency used to obtain locally a high-accuracy measurement of the longitudinal position.
The intensity dynamics should further include a rotating structure with very low frequency from which it is possible to determine the global distance and discriminate between different fast-varying periods. 
Both frequencies need to be adjusted such that each period of the high-accuracy measurement can be distinguished from any other using the slow rotation pattern.
In the theoretical description, this idea can be implemented by adding a third term to the equation earlier introduced \eqref{eq:field2mo}, such that we obtain
\begin{eqnarray}
    u(r,\phi,z) &=& A_{\ell_1}(r)\exp{[i(k_{1z}z+\ell_1\varphi)]} \nonumber \\
    &~& + A_{\ell_2}(r)\exp{[i(k_{2z}z+\ell_2\varphi)]} \nonumber \\
    &~& + A_{\ell_3}(r)\exp{[i(k_{3z}z+\ell_3\varphi)]}.
    \label{eq:field3mo}
\end{eqnarray}
As can be seen, the electric field defined above contains three different contributions, each characterised by its spatial frequency $k_i$ and OAM $\ell_i$. 
If we now calculate the intensity distribution generated by such a field, we will have, together with the contributions of the single terms in \eqref{eq:field3mo} -- i.e., terms proportional to $|A_{\ell_k}|^2$ -- also all the possible interference terms between the three beams composing the field above. 
We can therefore write, neglecting the $z$- independent terms, which amount only to an overall normalisation factor (supplemental material of \cite{vetter2014generalized}),
\begin{eqnarray}
    I &\propto& \cos^2{[\Delta k_{1,2} z + \Delta \ell_{1,2} \varphi]} + \cos^2{[\Delta k_{1,3} z + \Delta \ell_{1,3} \varphi]} \nonumber \\
    &~& + \cos^2{[\Delta k_{2,3} z + \Delta \ell_{2,3} \varphi]},
\end{eqnarray}
where we labelled $\Delta k_{i,j}= (k_{iz} - k_{jz})/2$ and $\Delta \ell_{i,j} = (\ell_i - \ell_j)/2$ as the pairwise differences between the wave vectors and OAM values of the three fields. 
By choosing $\Delta \ell_{2,3}=0$, i.e., $\ell_2=\ell_3$, the angular dependence of the propagation dynamics of the last term vanishes. 
Hence, we obtain the required light field, whose structure rotates with only two rotation frequencies at the same time.

\subsection{Experimental implementation}

In an experiment, it is convenient to realize these radially accelerating beams in the framework of Bessel beams, i.e., $A_\ell(r)=J_\ell(rk_r)$ \cite{ornigotti2018vector,rop2012measuring}. 
For Bessel beams the longitudinal wavevector $k_{z}$ can straightforwardly controlled by adjusting its radial counterpart $k_r$. 
Both quantities are related through 
\begin{equation}
    k_z=\sqrt{k^2-k_r^2},
\end{equation}
where $k=2\pi/\lambda$ labels the wavenumber and $\lambda$ corresponds to the wavelength of the utilized light field.

As the angular spectrum of a Bessel beam forms a ring in $k$-space, such beams are relatively simple to generate in the laboratory. 
By modulating an incoming light field to have a ring-shaped amplitude at one plane ($k$-space), we can transform the light into a Bessel beam (real space) by implementing an optical Fourier transform using a properly placed lens. 
The lens is placed one focal distance $f$ away from the initial modulation plane, i.e., in the back focal plane of the lens, such that at around another focal distance behind the lens a Fourier transform leads to a Bessel beam with a well-defined longitudinal wave vector $k_{z}$. 
In the modulation plane is a phase-only spatial light modulator (SLM) whose screen displays a holographic pattern, generated by a MATLAB script, that modulates the amplitude and phase of the beam into the required ring shape \cite{rosales2017shape}.
Depending on the radius $r$ of the ring in the modulation plane, the Bessel beam with longitudinal wave vector
\begin{equation}
    k_z = \frac{2\pi}{\lambda} \cos{\left(\frac{r}{f}\right)}
\end{equation}
will be obtained, which follows from simple geometric arguments \cite{vasilyeu2009generating,rop2012measuring}.

A radially self-accelerating beam, such as the one described above, follows from simply modulating the light field to have two (or more) rings with different radii $r_i$ and individual OAM values $\ell_i$.
The spiraling around the optical axis can thus be tuned by changing the radii of the two rings, leading to a wave vector difference
\begin{eqnarray}
    \Delta k_{i,j} &=& (k_{iz} - k_{jz})/2\nonumber\\
            &=& \frac{\pi}{\lambda} \left[\cos{\left(\frac{r_i}{f} \right)}-\cos{\left(\frac{r_j}{f} \right)} \right].
\end{eqnarray}
As discussed earlier, to achieve the best possible longitudinal accuracy, we aim at generating a structure that contains both a very high and a very low rotation frequency at the same time. 
To realise such a rotating structure, the light field generated by the SLM needs to contain three Bessel beam components, so that their mutual coupling gives rise to a fast-rotating and a slow-rotating self-accelerating beam.
The former, for example, results from Bessel beam components 1 and 2, whose difference in radii should be as large as the optical system allows, while the OAM value differs by only a single quanta, i.e., $\Delta \ell=1$.
As one of the rings, ring 2, necessarily has to have a large radius $r_2$ leading to a large difference in longitudinal wave vectors $\Delta k_{1,2}$, the interfering light field and, thus, the rotating structure will be strongly confined at a small transverse region around the optical axis.

The slow-rotating part, on the other hand, results from Bessel components 1 and 3, so the radii of the two rings should be very similar to obtain a small difference in wave vectors $\Delta k_{1,3}$.
However, the difference should also be large enough that the resulting rotation frequency allows us to discriminate between the repeating periods of the fast-oscillating signal.
If these two rings are chosen to be similar in radius but much smaller than ring 2, the rotating light field in the Fourier plane of the lens, i.e., the slowly rotating structure, will cover not only the area in close proximity of the optical axis but also the outer region. 
Note that examples of the ring-shaped modulation patterns and the resulting propagation dynamics of the spiraling beams can be found in Figures \ref{fig:sim1} and \ref{fig:sim2} in later sections. 
This difference in the radial intensity compared to the fast-rotating structure enables us to discriminate the two differently varying patterns by observing the intensity in different radial regions. 
In the simplest case, these regions might be defined by a single transverse location where the intensity is evaluated. 
However, higher experimental accuracy can be obtained using two quadrant detectors, one for each rotating structure, which evaluate the differences between opposing quadrants to increase the signal-to-noise ratio.
As such detectors can work with tens of nanoseconds of rise time, the proposed method might also find applications in high-speed longitudinal position measurements.

Another important aspect is the overall distance over which the spiraling intensity can be observed.
Here, the ultimate limit is given by the width of the ring in the modulation plane that generates the Bessel beam. 
The narrower the ring, the longer will the spiraling beam survive, thus allowing a measurement for longer distances. 
On the contrary, the wider the ring width is, the shorter will be the self-accelerating beam in the focal region, thus allowing measurements over only a very short distance. 
Obviously, in applications the ring width is determined by the resolution of the modulation device, which is in our case the SLM (see below), as well as the minimum amount of light required to detect the rotating intensity patterns. 
Using high-resolution generation methods, self-accelerating beams over a distance of 70~mm have been demonstrated already \cite{vetter2019realization}.

We further note that, in principle, one can also tune the rotation frequency by adjusting the difference in the OAM values $\Delta l$ of the two constituent beams, as can be seen in formula \eqref{eq:rot}. 
However, by doing so, one should keep in mind that large differences in OAM result in beams of more complex angular structures, which might also require a detection system that is able to resolve those structures.
Moreover, if the two constituents differ by many OAM quanta, they also show a significant difference in their OAM-induced divergence of the beam \cite{padgett2015divergence}. 
As this difference also leads to a fast decrease in their spatial overlap, the region over which the interference and, thus, the rotation can be observed is reduced.
Hence, improving the accuracy as well as the distance over which it can be measured is preferably done by tuning the rotation frequency through adjustments to the longitudinal wave vector difference $\Delta k$.

\section{Results} 

\subsection{High-accuracy measurements}
\begin{figure}[h]
    \centering
    \includegraphics[width=0.5\textwidth]{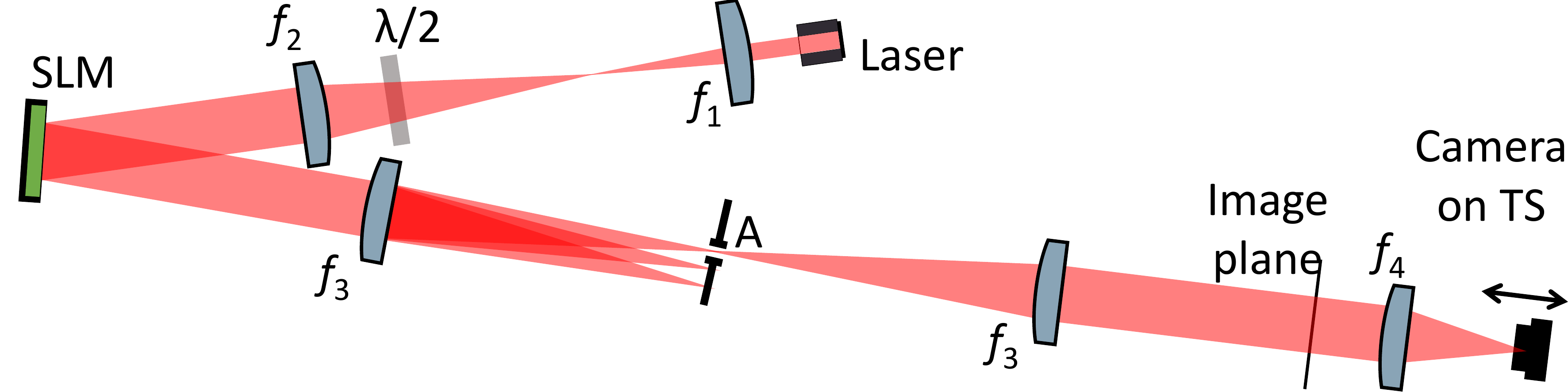}
    \caption{Sketch of the experimental setup used in simulation and experiment. A laser beam with a wavelength of 780~nm is enlarged to a wide radius using a telescope (lenses $f_1$ and $f_2$). For maximum efficiency, the beam's polarization is controlled by a half-wave plate $\lambda/2$ such that it aligns with the orientation of the diffraction grating on the SLM, which modulates the diffracted light to have a shape with a central circle and ring(s) and imparts angular momentum to the ring(s). The beam is filtered through a 4f optical system (lenses $f_3$) that extracts the first diffraction order, thereby removing the un-modulated light. Beyond the image plane, the structured beam is focused by another lens $f_4$. The resulting spiraling structure is imaged by a camera on a translation stage (TS). } 
    \label{fig:ExpSetup}
\end{figure}

As a first task, we investigated the largest possible rotation frequency and, thus, the highest possible accuracy.
Before testing the theory in the laboratory, we performed the so-called split-step propagation method\cite{poon2017engineering} to simulate the entire setup. 
A sketch of the setup can be seen in Fig. \ref{fig:ExpSetup}.

During the first set of simulations, the main aim was to verify the idea and find the fastest rotation of the angular modes that was still within the limitations of our experimental system. 
These limitations were mainly the aperture of our optics (1-inch); the pixel resolution of our modulating device, a phase-only SLM (Holoeye, Pluto-2.1-NIR-011 LCOS, 1920$\times$1080 pixels, 8~$\mu$m pixel size); and the pixels of the detection system, a camera (ZWO ASI120MM Mini, 1280$\times$960 pixels, 3.75~$\mu$m pixel size).

In order to achieve the best matching between simulation and experiment, we included in the simulation the spatial resolution of our modulation and detection plane using the values of our laboratory system.
In particular, we set the pixel size of the simulation to 1.875~$\mu$m, which is half the size of the physical pixels of our camera used in the experiment.

\begin{figure}[h]
    \centering
    \includegraphics[width=0.5\textwidth]{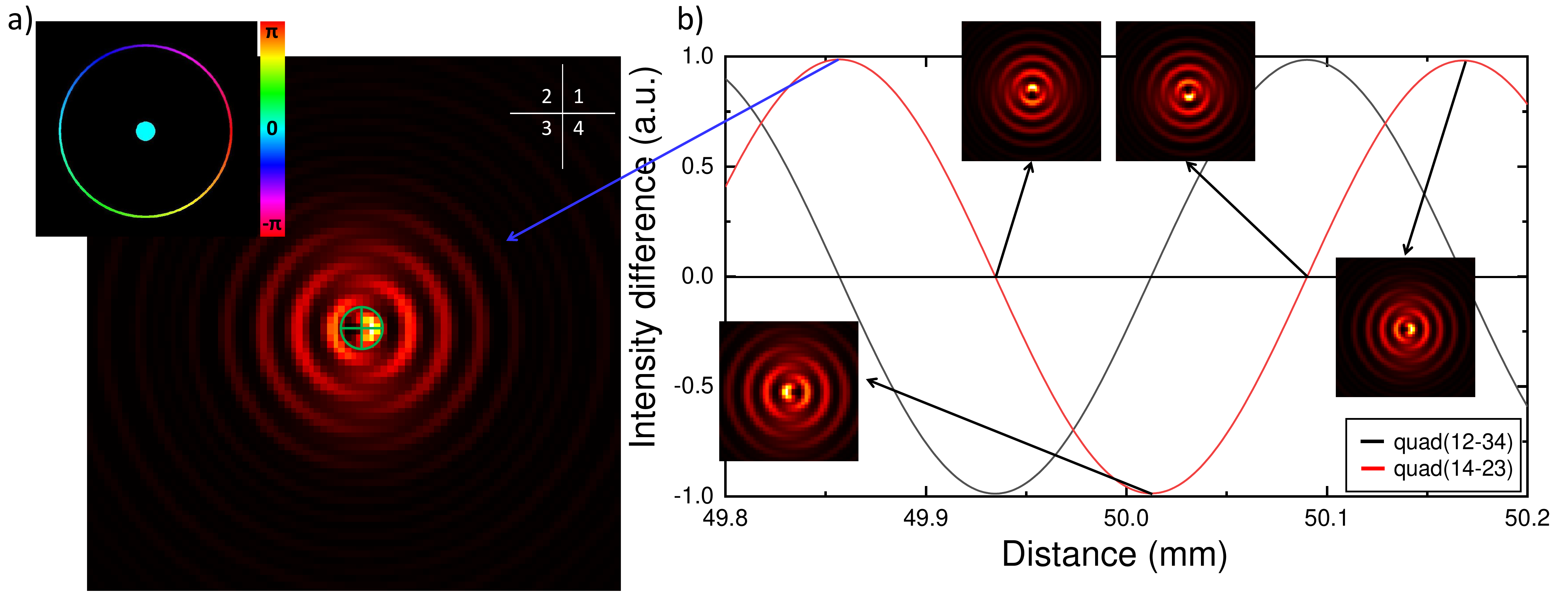}
    \caption{Simulation for high-accuracy measurements. 
    a) In the inset, the modulation pattern is shown to generate the spiraling structure. Its brightness corresponds to the amplitude of the light, and the color depicts the phase of the modulation. For clarity only the modulation is shown, not the holographic pattern required in the experiment.
    The green lines depict the region utilized to emulate a quadrant detector.
    b) The simulated angular intensity changes over propagation distance.
    Four exemplary intensity patterns are shown as insets to visualize the spiraling behaviour.
    Comparing the intensities found in 4 quadrants, shown in a), around the optical axis enables the determination of the longitudinal position.
    The intensity difference between quadrants 1,2 and 3,4 (2,3 and 1,4) leads to a sinusoidal curve shown in black (red) with a period of 311.2~$\mu$m.
    Using both fast-varying curves the longitudinal positions can be unambiguously determined over one half period.} 
    \label{fig:sim1}
\end{figure}

At first, we started by simulating a beam that only rotates with a single frequency, i.e., a beam consisting of two Bessel components, as described in equation \eqref{eq:field2mo}.
To achieve the largest difference in their longitudinal wave vectors, we generated two Bessel beam components from two ring-shaped patterns of very different radii in the Fourier domain.
The larger ring size was mainly limited by the screen size of the SLM as well as the optical apertures in our system.
The ring we utilized had an inner radius of $r_1$=3.5~mm, with a ring width of 0.1~mm, and imprinted an OAM value of $\ell=$1 onto the beam.
As the length over which the rotating intensity pattern exists is determined by the ring width, we chose the smallest possible width allowed by the resolution of the modulation device, i.e., the pixel size of the SLM. 
Utilizing an SLM with a pixel size of 8~$\mu$m, we chose a ring width of 100~$\mu$m in order to have around 11-13 pixels at any given angular position, which allowed an efficient generation using holographic methods. 
We chose the smaller ring to be a circular area of radius $r_0 = 0.4$~mm with a flat phase, i.e., an OAM value of $\ell=0$ (see Fig. \ref{fig:sim1}a).
The circular area was optimized to have a similar intensity as the ring, taking the Gaussian shape of our input light field, with a beam waist of 4.1~mm, into account.
This optimization was done because equal amplitudes of the two components result in a rotating pattern with improved visibility \cite{vetter2015optimization,ornigotti2018vector}, which in our task improves the accuracy.

Finally, the ring-shaped intensity patterns were transformed into a spiraling beam through an optical Fourier transform using a lens of 50~mm focal length.
The focal length was mainly limited by the pixel size of the camera, as shorter focal lengths lead to beams of smaller extent in the focus. 
We note that a magnification system might be used to circumvent this constraint if shorter focal length lenses are required.

In order to determine the propagation distance from the change in angular intensity, we simulated the intensity in a 300$\times$300-pixel region centered on the optical axis during propagation.
We then registered the intensity within a circular region of radius 3.75~$\mu$m around the optical axis.
To emulate a quadrant detector we summed up the registered intensity of each of the four quadrants and evaluated the difference between the intensities of quadrants 1,2 and 3,4, i.e., the upper and lower halves (see Fig. \ref{fig:sim1}a).
We obtained a sinusoidal curve with a period of 311.2~$\mu$m, as shown in Fig. \ref{fig:sim1}b, which matches the value of 311~$\mu$m expected from theory.
In addition, if one evaluates the difference in intensity between quadrants 2,3 and 1,4, i.e., the left and right halves, it is possible to shift the steepest slope of the sinusoidal curve by a quarter of a period (an effective phase shift of $\pi/2$).
When using both signals, it is possible to achieve the same longitudinal accuracy at the positions where the shallow slope of one signal would cause a significant decrease in accuracy (see Fig. \ref{fig:sim1}b).

After having verified the method in simulations, we turned to the experimental implementation to determine the actual accuracy limits of our system due to experimental imperfections and errors. 
In the experiment, sketched in Fig. \ref{fig:ExpSetup}, we used a fiber-coupled single-frequency Toptica laser (DLpro) at a wavelength of 780~nm. 
We enlarged the laser beam using a telescope system to a beam waist radius of approximately 4.1~mm, such that it illuminated the whole screen of the SLM.
Via reflection off the SLM screen, we modulated the beam to have the multiple-ring shaped intensity structure with the dimensions described above.
As the SLM is a liquid-crystal phase-only modulation device \cite{zhang2014LCOS}, we used holographic modulation techniques to perform a complex amplitude modulation. 
This is done by displaying the diffractive holographic pattern only at the ring-shaped regions where we wanted to obtain a light field \cite{rosales2017shape}. 
Through filtering only the first diffraction order using an aperture in a Fourier plane of a 4\textit{f}-system, we not only imprint the required phase but also carve out the required intensity structure.
At the image plane, we obtained the required ring-shaped amplitude that leads to a spiraling beam after a Fourier transformation, as described earlier.
Analogous to the simulations, we performed this Fourier transformation using another lens with a focal distance of 50~mm.

To record the spiraling structure, we placed the camera on a high-accuracy motorized translation stage, which was scanned over 1~mm in steps of 1~$\mu$m. 
The recording and translation of the camera were automated by interfacing the camera and translation stage with LabVIEW. 
At each position, we recorded 50 frames, from which we obtained the average and standard deviation of the angular intensity along the optical axis.
Analogous to the simulation, we emulated a quadrant detector by registering the intensity difference between different quadrants illuminated by the beam.
As the beam waist in the focal plane is very small, we only used 4 pixels of size 3.75~$\mu$m, each corresponding to one quadrant.

\begin{figure}[h]
    \centering
    \includegraphics[width=0.45\textwidth]{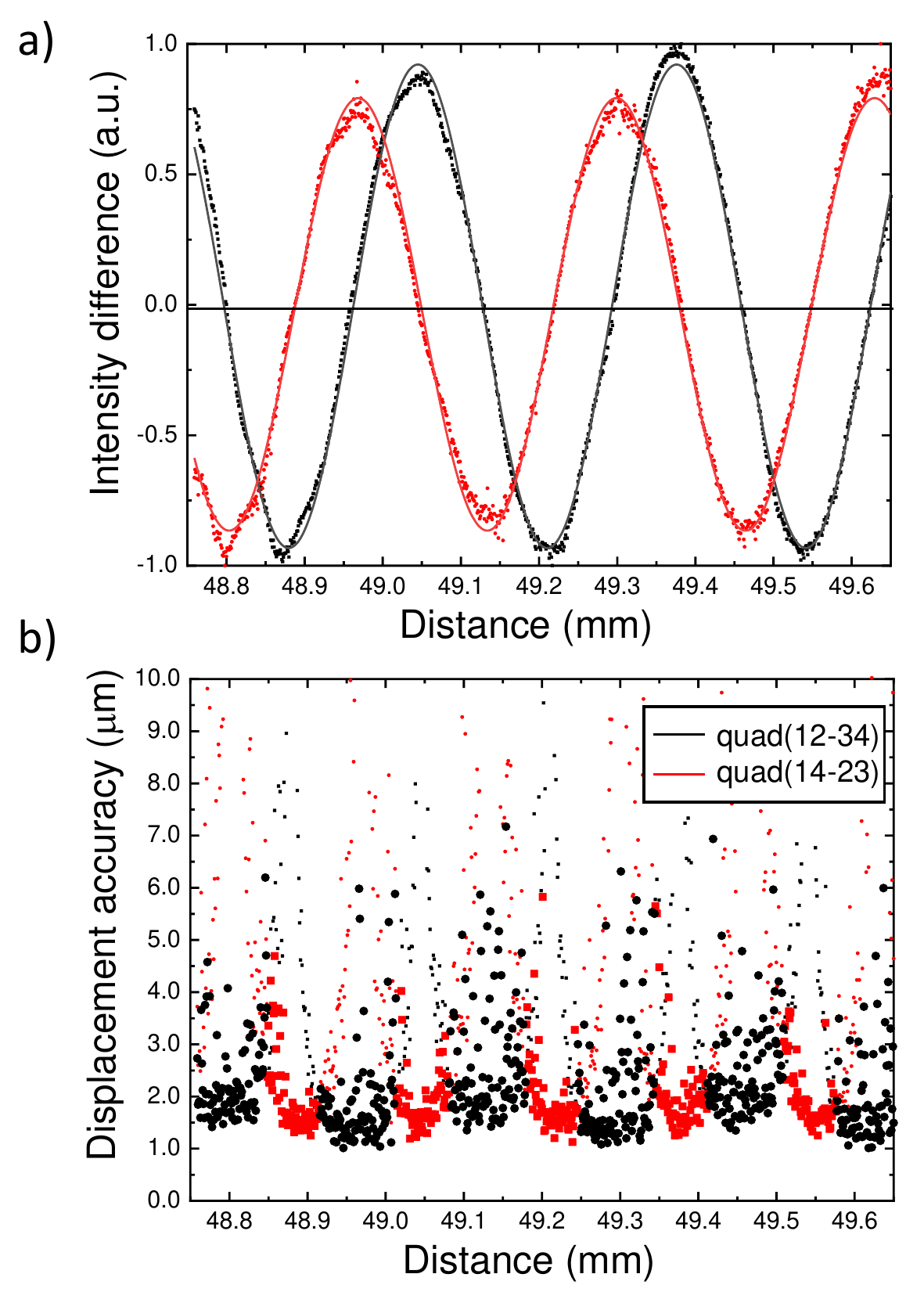}
    \caption{Experimental high-accuracy distance measurements.
    a) The scatter plot of intensity differences between quadrants 1,2 and 3,4 (2,3 and 1,4) are shown in black (red). 
    The intensities are obtained by averaging 50 images for each position.
    The solid lines show the corresponding sinusoidal fits to the measurements.
    b) The displacement accuracy analysis is obtained using the fitted sine functions and the error propagation method from the standard deviation of intensity fluctuations at each position. 
    We find a best accuracy of 1-2~$\mu$m at the points of steepest slope. 
    For maintaining the minimum error, one must hop between the steep slopes of the red and black curves, which leads to obtaining the accuracy of 2.2$\pm$0.9~$\mu$m over the full range (larger symbols).}
    \label{fig:RelPosmeas}
\end{figure}

The resulting variations in intensity when comparing quadrants 1,2 and 3,4 are shown in Fig. \ref{fig:RelPosmeas}a, demonstrating a period of about 332.6$\pm$0.3~$\mu$m. 
This result matches the expected period from theory and simulation, with the small discrepancy attributed to experimental imperfections such as finite resolution and misalignment.
Using the standard deviation of the intensity at each position as the experimentally determined error and taking the known sinusoidal curve into account, we determined the longitudinal accuracy or displacement accuracy through error propagation.
Hence, we define the displacement accuracy as the minimal displacement for which the errors of two data point do not overlap, i.e., the two data points that can be discriminated with a 1-$\sigma$ confidence interval.
We found that at the steep slope of the curve, two different longitudinal positions that are 1 to 2~$\mu$m apart can still be resolved with one standard deviation significance (see Fig. \ref{fig:RelPosmeas}b).
As expected, the accuracy decreases dramatically around the extremal regions of the sinusoidal curve.
However, as mentioned earlier, evaluating the difference in intensity between the left and right halves (quadrants 2,3 and 1,4) results in a periodic signal that is shifted by a quarter of a period relative to the signal of the upper and lower halves.
We can therefore always refer to a fast-varying signal regardless of position, so the strong reduction of accuracy due to slow intensity variations at the extremal points of the sinusoidal curve is circumvented. 
When switching between the two signals such that the quarter period with the steepest slope is always used for a given longitudinal region, we obtain an average accuracy of 2.2$\pm$0.9~$\mu$m over the whole scanning range.

\subsection{Long range measurements}
In order to overcome the ambiguity between multiple fast-varying periods, we then studied a beam that spirals with two components simultaneously: a slow one and a fast one.
The former can be used to determine the coarse position, while the latter can be used to obtain the longitudinal distance with high accuracy.
Again, we first investigated the method in simulation before implementing the scheme in the laboratory.

\begin{figure}[h]
    \centering
    \includegraphics[width=0.5\textwidth]{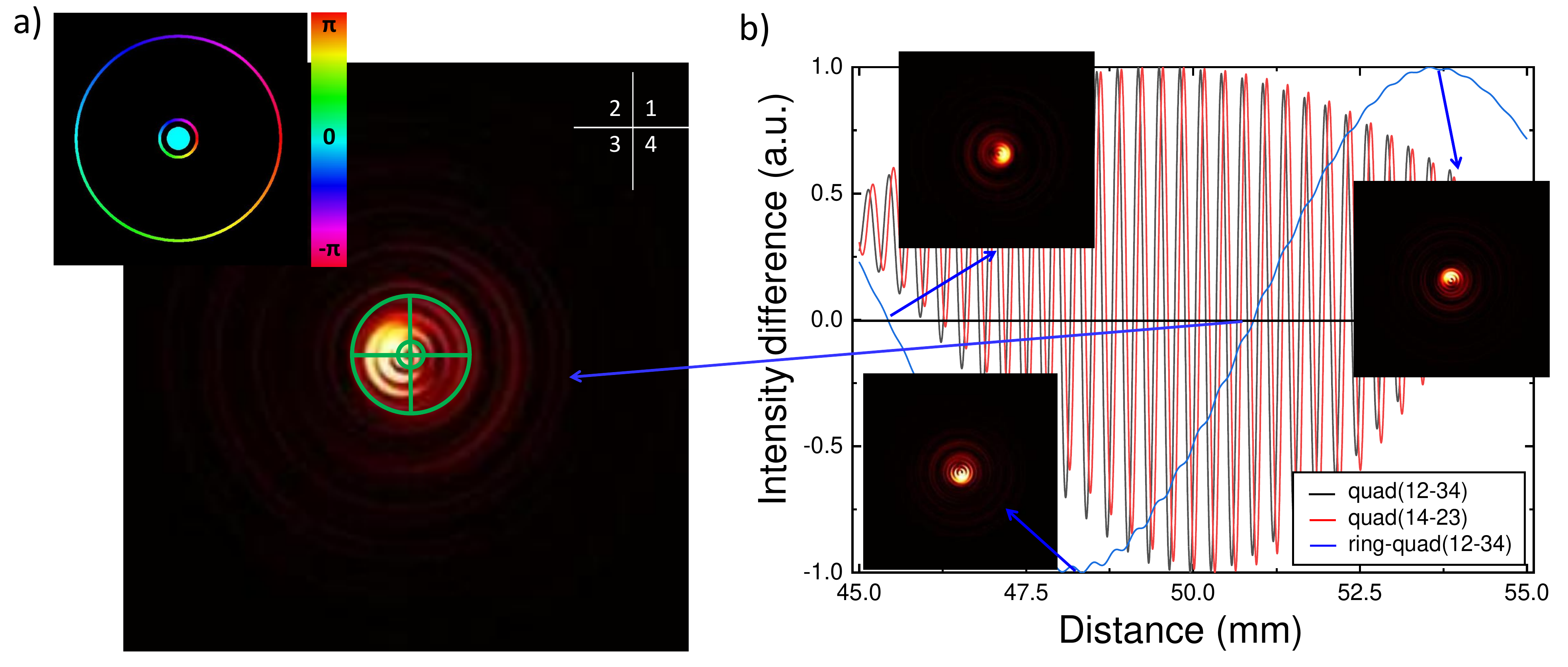}
    \caption{Simulation for distance measurements over a longer range. 
    a) An intensity structure spiraling simultaneously with two frequencies is generated using the modulation pattern shown in the inset. As in Fig. \ref{fig:sim1}, only the complex modulation is shown and the brightness depicts the amplitude and the color depicts the phase. 
    The simulated beam close the focal region shows the small off-axis intensity, that is spiraling fast as before, as well as larger regions of increased intensity, which is simultaneously spiraling slower around the optical axis.
    The regions used to emulate the quadrant detectors are depicted with green lines.
    b) The angular intensity variation found in the inner and out regions around the optical axis over propagation distance. Comparing the intensity differences found in four quadrants in the inner region (1,2-3,4 in black and 2,3-1,4 in red) around the optical axis leads to a fast-varying pattern used as a local high-accuracy measurement analog as before. Comparing the intensity difference in the outer region (ring-quad, 1,2-3,4), however, leads to a slow-varying function (blue line) with a periodicity of 12.63~mm, which will give information about the global position such that the high-accuracy distance measurement can be extended over multiple periods, i.e., a longer region. 
    Small oscillations in the slow-varying curve are due to an inevitable cross-talk from the strong fast-oscillating signal.} 
    \label{fig:sim2}
\end{figure}

As discussed in the theory section, such a dual-frequency beam can be generated by adding an additional ring to the modulation pattern. 
In our realization, we used an additional ring that was slightly bigger than the inner circular area, with an inner ring radius of $r_1$ = 0.6~mm and a width of 100~$\mu$m.
The modulation pattern to generate such a beam can be found in Fig. \ref{fig:sim2}a.
As before, the ring width was limited by the pixel size of the modulating device used in the subsequent experiment.
The additional ring results in a beam having a second, much smaller rotation frequency in the intensity, as described by equation \eqref{eq:field3mo}.
To prevent a third rotation frequency from appearing in the intensity pattern, we imprinted the additional ring with an OAM value of $\ell=1$, such that the pairwise interference only appears between the circular central area and each of the rings. 
Again, the width of the additional ring was chosen such that all three beam components were similar in amplitude, thus obtaining a high-visibility structure \cite{vetter2015optimization}.
The rest of the simulation remained the same as before.

\begin{figure}[h]
    \centering
    \includegraphics[width=0.45\textwidth]{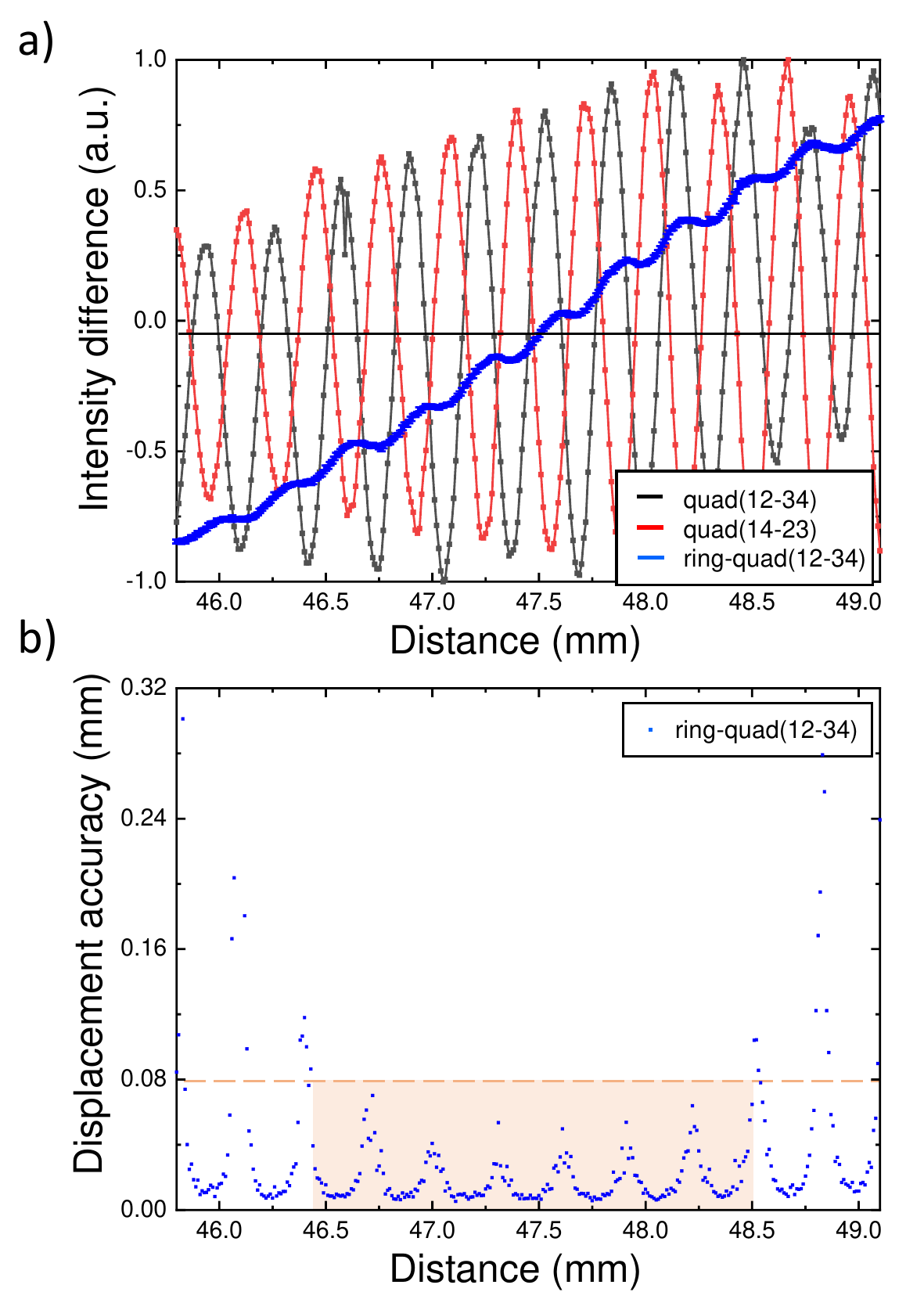}
    \caption{High-accuracy distance measurements over a long range. 
    a) The connected-scatter plot of intensity differences between the inner quadrants (red and black) and the ring quadrant (blue) as shown in Fig. \ref{fig:sim2}a.
    At every position, 25 images were recorded.
    b) Displacement accuracy analysis for the slow-varying curve (blue), again obtained using the fitted function from a), the standard deviation of the intensity fluctuations at each position, and error propagation.
    To distinguish the different steep slopes of the fast-oscillating signals, an accuracy of around 80~$\mu$m is required (dashed line), which we achieved over a region of about 2~mm (orange shaded region).
    } \label{fig:ActPosmeas}
\end{figure}

Because the components of the light field that rotate slower are stemming from the components on the inner parts of the generation hologram, the resulting interference is distributed over a larger region around the optical axis (see Fig. \ref{fig:sim2}a and insets in b).
In other words, the angular intensity found in regions of larger radii is strongly determined by the slow-varying pattern. 
This spatial separation enables a simultaneous measurement of both variations: 
the fast-spiraling part close to the optical axis and the slow-spiraling part further away from the beam center. 
As before, we used the small circular region with a radius of 3.75~$\mu$m around the optical axis to determine the fast variation.
The slow-varying signal was obtained by recording the intensity in a circular, ring-shaped region with an inner radius of 7.5~$\mu$m and a width of 22.5~$\mu$m.
Upon propagation, the intensity follows a fast as well as slow-varying sinusoidal curve when the differences between the quadrants are evaluated (see Fig. \ref{fig:sim2}b).
As expected, the fast oscillation period is again 311.2~$\mu$m, while the slow-varying signal has a periodicity of around 12.63~mm. 
While the fast-oscillating period again matches nicely with the theoretical value of 311~$\mu$m, we find a bigger discrepancy of the slow-oscillating signal to the theoretically expected periodicity of 10.2~mm. 
We attribute the latter to the fact that we only obtained only one full fringe, which also shows an additional modulation.
The additional slow-varying signal now allows a discrimination between the different fast-oscillating periods and, thus, a global determination of the longitudinal position.
In other words, after a first gauging of slope of the slow-varying curve with respect to a coarse distance measure, the fast-oscillating signal can then be used to determine the distance with high accuracy.
However, a closer inspection of the slow-varying curve shows that it is also slightly modulated by the fast-oscillating signal, as the two differently spiraling fields are not complete separable. 
While this leads to an increase of the slope in some regions, it also flattens the curve whenever the fast-varying modulation counteracts the change of the slow-oscillating curve. 
Obviously, in these regions the accuracy of the slow-varying signal will be reduced.
Thus, an important experimental question is to determine if the detrimental effect of this additional modulation is small enough to allow a discrimination between the different fringes using our quadrant measurement technique: in other words, over what range is the displacement accuracy of the slow-varying signal less than $1/4$ of the period of the fast-varying signal?

Apart from changing the modulation pattern at the SLM, the rest of the experimental setup remained the same as before.
However, this time we scanned over a range of 10~mm with a step size of 10~$\mu$m and recorded 25 image frames of the spiraling structure at each position.
In the data analysis, we again used the four pixels around the optical axis (2$\times$2 pixel array) as a quadrant detector to obtain the fast-oscillating signal. 
Additionally, to measure the slow-varying signal, we measured the intensity differences between the upper and lower quadrants (1,2 and 3,4) in a disk-shaped region with a radius of 30~$\mu$m around the optical axis, excluding the four inner pixels used to determine the fast-oscillating signal. 
As can be seen in Fig. \ref{fig:ActPosmeas}a, we find a slow-varying curve with a periodicity of 11.29$\pm$0.03~mm and fast-oscillating fringes with periods of 312.9$\pm$0.4~$\mu$m, as expected.
We also obtain, analogous to the simulations, the additional high-frequency modulation of the slow-varying signal.
To evaluate the displacement accuracy, we first determine the function that describes the slow-varying curve, for which we use a sum of two sinusoidal functions whose amplitude and periodicity we obtain from fits.
Using this function; experimentally obtained errors given by the standard deviation of the measured intensities; and error propagation, we find that we obtain the required accuracy of less than than 80~$\mu$m ($\sim1/4$ of the period of the fast curve) over a range of more than 7 fast-oscillating fringes, or around 2~mm. 
Most displacement accuracies for the slow-varying curve in this region are as low as 10-20~$\mu$m (see Fig. \ref{fig:ActPosmeas}b).
This accuracy is enough to distinguish the different fast-oscillating fringes and the corresponding regions having a steep slope, thereby demonstrating an accuracy of around 2~$\mu$m over the full region, i.e., three orders of magnitude or 2~mm.
We note that in principle our method does not require an initial calibration, as the longitudinal-varying angular intensities can be theoretically obtained from the beam dimensions as well as the focusing lens.
However, in experiments the system might be initially characterized once, such that a camera (or quadrant detector) placed anywhere within the possible measurement range can be used to determine an absolute position without rescanning the entire translation region.

\section{Conclusion} \label{sec:conc}
We have demonstrated that radially self-accelerating beams can be used to determine the distance with respect to a reference with an accuracy of 2~$\mu$m over three orders of magnitude. 
The main benefit of our technique is its simplicity, as only a single beam having the appropriate structure and two quadrant detectors are required.
Apart from this strong benefit, there is also an important precaution worth mentioning.
The technique requires a very accurate alignment of the beam's optical axis with the center of the detector such that the recorded intensity does not move transversely while the detector is translated. 
Even a single pixel of transverse displacement can cause large discontinuous jumps in the measured intensity differences.
However, once alignment has been ensured, the measurement is stable and only requires minimal post-processing (after having gauged the system), such that very high read-out speeds on the order of nano-seconds might be feasible.
The obtained accuracy of the distance measurements can be further improved by using stronger focusing optics and custom-fabricated optics for beam generation, as well as an additional imaging system to lift the limitation of the finite resolution of the detection system.
In addition, it might be interesting to consider more complex spiraling structures, such as an accelerated rotation \cite{schulze2015accelerated}, which should further increase the accuracy over small regions at the cost of the accuracy elsewhere.
Finally, we hope to stimulate further research into applications that benefit from the propagation-dependent intensity variations of radially self-accelerating beams.

\section*{Acknowledgments} \label{sec:ack}
SP, SP, MO, and RF acknowledge the support from Academy of Finland through the Competitive Funding to Strengthen University Research Profiles (decision 301820) and the Photonics Research and Innovation Flagship (PREIN - decision 320165). RF also acknowledges support from Academy of Finland through the Academy Research Fellowship (decision 332399).

\section*{Disclosures}
The authors declare that there are no conflicts of interest related to this article.

\end{document}